\documentclass{WileyMSP-template}
\usepackage{cite}
\usepackage{amsmath}
\usepackage[utf8]{inputenc}
\usepackage[T1]{fontenc}
\usepackage{textcomp}
\usepackage{caption}
\usepackage{bm}
\usepackage{float}
\usepackage{threeparttable}
\DeclareUnicodeCharacter{2212}{-}

\usepackage[]{lineno}
\usepackage{xcolor}

\begin{document}

This work has been submitted to Wiley for possible publication. Copyright may be transferred without notice, after which this version may no longer be accessible.

\newpage

\pagestyle{fancy}
\rhead{\includegraphics[width=2.5cm]{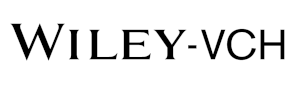}}
\title{Reconfigurable Wearable Antenna for 5G Applications using Nematic Liquid Crystals}
\maketitle

\author{Yuanjie Xia}
\author{Mengyao Yuan}
\author{Alexandra Dobrea}
\author{Chong Li}
\author{Hadi Heidari} 
\author{Nigel Mottram}
\author{Rami Ghannam*}

\begin{affiliations}
Mr Y. Xia, Ms M. Yuan, Ms A. Dobrea, Dr C. Li, Prof. H. Heidari, Dr R. Ghannam\\
James Watt School of Engineering, University of Glasgow, Glasgow G12 8QQ, UK.

\vspace{5mm}
Prof. Nigel Mottram\\
School of Mathematics and Statistics, University of Glasgow, Glasgow G12 8QQ, UK.\\

\vspace{5mm}
Email Address: rami.ghannam@glasgow.ac.uk
\end{affiliations}



\begin{abstract}
 
The antenna is one of the key building blocks of many wearable electronic device, and its functions include wireless communications, energy harvesting and radiative wireless power transfer (WPT). In an effort to realise lightweight, autonomous and battery-less wearable devices, we demonstrate a reconfigurable antenna design for 5G wearable applications that require ultra-low driving voltages ($0.4$-$0.6\,$V) and operate over a high frequency range ($3.3$-$3.8\,$GHz). For smart glasses application, previous antenna designs were `fixed' and mounted on the eyeglass frame itself. Here, we demonstrate a reconfigurable design that could be achieved on the lens itself, using an anisotropic liquid crystal (LC) material.  We demonstrate how LC alignment and electric field patterns strongly influence the tuning capabilities of these antennas in the gigahertz range and present a smart, reconfigurable spiral antenna system with a LC substrate.

\end{abstract}

\section{Introduction}
The electronic antenna is a circuit element that transforms a guided wave in transmission lines into a free-space wave and vice versa. Antennas are essential components for wireless communication systems and are widely used in both wearable and implantable devices. The two main applications of antennas in the literature are energy harvesting \cite{lingley2011single} and data transmission \cite{chen2017warpage}. In this manuscript, we demonstrate how an antenna's frequency can be tuned using Liquid Crystal (LC) materials, making it appealing for a range of energy harvesting and telecommunications applications.

Multi-functional wearable devices require both a wide frequency bandwidth and a miniaturized scale. For energy harvesting applications, a tunable frequency bandwidth device enables multiple radio frequency (RF) wavelengths in free space to be harvested. Similarly, for communications applications, a tunable antenna enables wearable devices to transmit and receive broadband information \cite{kim2021design}, and helps reduce mismatch and noisy interference when communicating with other devices \cite{martin2003patch}. In addition, and in comparison with traditional antennas, a tunable antenna can establish communication with multiple devices with different frequencies, which could play an important role in Multiple In Multiple Out (MIMO) networks \cite{jakoby2020microwave}, making dynamically tunable antenna front ends a necessity in the near future. Consequently, we demonstrate a tunable antenna design that covers the frequency range of communications systems and consumes minimum energy in a limited area. Such an antenna design could be mounted on wearable or implantable devices to enable both communications and energy harvesting capabilities.

There are many tunable antenna strategies reported in the literature. For example, the method demonstrated by Huang \textit{et al.} involved tuning the impedance of the matching network to change the reflection coefficient of frequency \cite{huang2008electrically}. Other previous studies attempted to achieve electronic reconfigurability of the reflective arrays using varactor diodes \cite{hum2005realizing}. Other methods involved using reconfigurable substrate materials, such as controllable dielectrics \cite{bildik2011reconfigurable,perez2015design} and Barium Strontium Titanate \cite{romanofsky2000k}. However, the complexity of these approaches are not optimal when targeting wearable devices. In terms of LC-tunable antennas, A. Gaebler \textit{et al.} devised a novel tunable antenna for millimeter wave applications \cite{gaebler2009liquid} using a commercial nematic liquid crystal (NLC) mixture, K15. According to their results, the bandwidth could be tuned from $33.25\,$GHz to $37.1\,$GHz with a driving voltage up to 20 V. Obviously, this driving voltage is impractical for a wearable application, and the purpose of this paper is to demonstrate that such a tunable wavelength range can be achieved using a much smaller driving voltage, making use of the previously-demonstrated results showing that LCs demonstrate excellent RF performance in the gigahertz range and that the frequency range and tunable bandwidth can be expanded using different LC mixture materials \cite{park2012evaluating, yang2018electrically}.

As utilized in the above-mentioned papers, and in the present work, the key benefit of LC-based antennas is their ability to exhibit analogue tuning of properties such as their permittivity and refractive index. This ability stems from the anisotropic nature of the liquid crystal phase which, at the molecular level, consists of anisotropic molecules which may be, for instance, rod-shaped or disc-shaped. This molecular shape anisotropy results in orientational, and sometimes positional, order, which then translates into a bulk phase anisotropy. This orientational ordering of the molecules means that the `director', the ensemble average direction of LC molecular orientation, can be defined \cite{mottram2014introduction}. These anisotropic properties allow LCs to be manipulated through a balance of forces and torques due to an applied electric field and elastic effects, yielding a reconfigurable design where the bandwidth of the antenna can be easily and accurately tuned \cite{shields2020control,jones2021vertically}. In the present work we will consider those liquid crystals which exhibit orientational anisotropy but no positional anisotropiy, termed `nematic' liquid crystals (NLCs).

\begin{figure}[H]
  \centering
  \includegraphics[width=18cm]{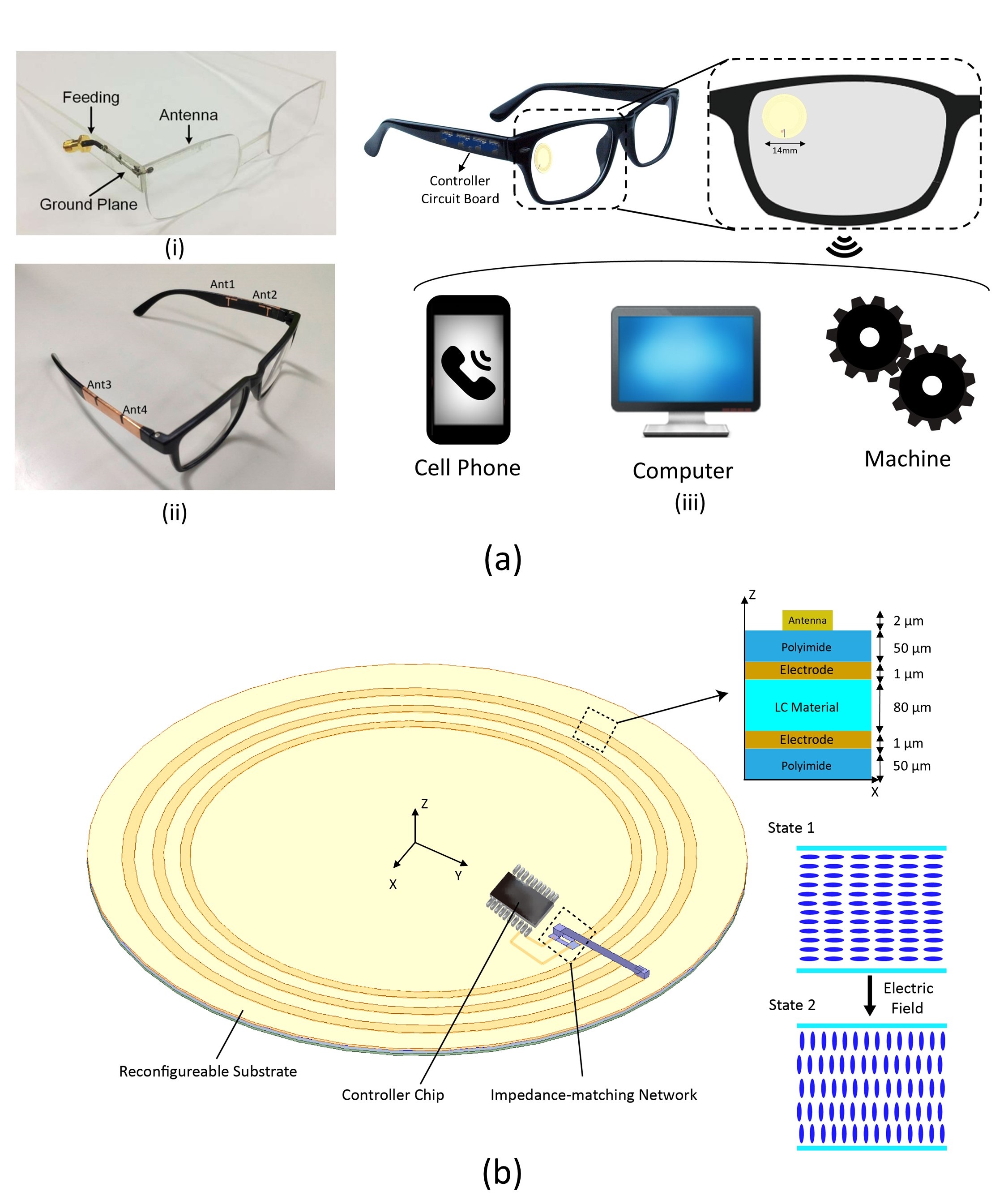}
  \caption{\textcolor{red}{(a) (i) and (ii) show previous antenna designs for smart glasses, where in (i) a transparent antenna is mounted on the lens \cite{hong2016transparent} and in (ii) a patch antenna is mounted on the  frame of the glasses \cite{chung20214}. (a) (iii) shows how the proposed antenna could fit on the lens of a smart glass in our design, as well as possible applications of this tunable antenna. (b) Schematic diagram of the tunable antenna and the structure of the reconfigurable layer.}}
  \label{Schematic_Diagram}
\end{figure}

Examples of antennas mounted on smart glasses are shown in Fig.~\ref{Schematic_Diagram}(a)(i) and (ii), \cite{chung20214,hong2016transparent}. In Fig.~\ref{Schematic_Diagram}(a)(i), the antenna was placed on the eye-glass frame. In addition to such an antenna being fixed, placement on the eye-glass frame results in a bulky glasses frame. In Fig.~\ref{Schematic_Diagram}(a)(ii), the antenna is instead placed at the far edge of the lens of glasses, which can reduce the size of the glasses frame without directly blocking the user's view. Here, we propose a tunable antenna that is integrated into `intelligent' glasses, where the antenna and the LC substrate are mounted on the glass lens, as shown in Fig.~\ref{Schematic_Diagram}(a)(iii). According to the literature, antennas mounted on the far edge of the lens can minimize the influence of vision obstruction  \cite{hong2016transparent}. Moreover, our proposed design consists of transparent materials to avoid blocking the user's view. Overall, the proposed antenna consists of a dielectric layer, an electrode layer and a LC layer, as shown in Fig.~\ref{Schematic_Diagram}(b). As previously mentioned, the LC director orientation is influenced by an applied electric field. Therefore, our hypothesis involves `tuning' the antenna's bandwidth by switching our LC directors from an initial planar state to a state determined by an applied voltage, which leads to a variation of the material's dielectric permittivity \cite{klasen1998calculation}. 

The key design parameters for this technology are therefore to fit the dimensions of typical eye glass lenses, have high transparency, operate at a low voltage and with a target frequency. With regard to transparency, previous transparent antenna designs have achieved average antenna transmittance of around 81\%, sufficiently high as to avoid impacting a user's vision \cite{hong2016transparent}, and in our design, we therefore aim for transparency higher than 80\%. Moreover, since high voltages are unacceptable for wearable devices, and where the power supply of smart electronics is usually around $3$ to $5\,$V \cite{jiang2017light} we aim for voltages lower than $5\,$V. As is mentioned in previous papers, the 5G network is will be used in the future development of most `Internet of Things' (IoT) applications \cite{borkar2016application}, so we aim to design and tune the antenna within a 5G frequency band.


In the present manuscript we demonstrate a `proof-of-principle' of using an LC for the design of an antenna with the above design requirements. We use LC and antenna simulations to demonstrate the feasibility of such a device. Our results illustrates that the proposed tunable antenna could be driving with $0.4$ to $0.6\,$V and cover the section of the 5G communication frequency range from $3.3$ to $3.8\,$GHz, which both more than adequately satisfy the design criteria for wearable applications. Apart from communication applications, the tunable antenna could also be applied to harvest energy from within the 5G frequency range. In comparison with previous designs, our design has a smaller footprint, thinner substrate material and lower driving voltage, which is more suitable for wearable applications.

\section{Methodology}
\subsection{Theoretical Framework}

As previously mentioned, the orientation of the NLC director can be altered through a balance of effects due to an applied electric (or magnetic) field and internal elasticity. The director configuration that minimises the total free energy, which comprises of electrostatic and elastic components, will be the one observed, and can be tuned by varying the applied voltage. These two effects are mathematically formulated through the electrostatic energy density $F_{es}$, which depends on the anisotropic dielectric properties of the material described in terms of a rank-2 dielectric permittivity tensor $\bm\epsilon$, and the bulk nematic energy density which consists of an elastic component $F_{el}$ and a thermotropic component $F_{th}$. Once the director profile that minimizes the free energy is determined, the optical phase profile of the LC material may then be calculated.

Although the free energy density components may be written in terms of  the director, ${\bf n}({\bf x})$, problems occur in such a models when high gradients of ${\bf n}({\bf x})$ are present, for instance near electrode edges. We therefore use the Landau de Gennes (LdG) model of a nematic liquid crystal, whereby the alignment of the molecules in the LC state is described in terms of both a degree of orientational order, $S({\bf x})$, and the preferred direction of orientation, ${\bf n}({\bf x})$. Each free energy component is then written in terms of the `${\bf Q}$-tensor', a rank-2 tensor, which takes the form: 
\begin{equation}
   {\bf Q} = S\left( {\bf n} \otimes {\bf n} − \dfrac{\rm \bf I}{3}\right),  
\end{equation}
where ${\rm \bf I}$ is the identity matrix. The total free energy ($\cal F$) of the nematic liquid crystal can be written as the volume integral of the sum of the elastic ($F_{el}$), thermotropic ($F_{th}$) and electrostatic energy ($F_{es}$) density functions:
\begin{equation}
   {\cal F}= \int_{V}^{} (F_{el} + F_{th} + F_{es})\,dV. 
\end{equation} 

The application of an electric field to a NLC induces dipoles, which may be caused by induced molecular dipoles and/or reorientation of enhanced existing molecular dipoles, which causes the director to align either parallel or perpendicular to the electric field, depending on the largest dielectric permittivity of the material. The preference for parallel or perpendicular alignment with an applied electric field is therefore determined by the dielectric anisotropy, $\Delta\epsilon= \epsilon_\parallel-\epsilon_\perp$, the difference in permittivities  parallel and perpendicular to the molecular axis \cite{collings2017introduction}. The dielectric displacement field due to these dielectric effects is then: 
\begin{align}
    {\bf D} &= \epsilon_0 {\bm \epsilon} {\bf E},
    \end{align}
where $\epsilon_0$ is the permittivity of free space and ${\bm \epsilon}$ is the dielectric permittivity tensor, which can be expressed in terms of the ${\bf Q}$-tensor as \cite{barberi2004electric}:
\begin{equation}
{\bm \epsilon} =  \frac{\Delta \epsilon}{S_{eq}} {\bf Q}+ \tilde{ \epsilon}\, {\rm \bf I},
\end{equation}
and where, $\tilde{\epsilon} = ({2\epsilon_\perp + \epsilon_\parallel })/{3} $ is the average permittivity and $S_{eq}$ is the order parameter in the bulk of the liquid crystal at which the experimental measurement of $\Delta\epsilon$ was undertaken. The resulting electrostatic energy density  of the NLC is:
\begin{equation}
F_{es} = -\frac{1}{2}{\bf D}.{\bf E} 
        =-\frac{1}{2}\epsilon_0 ({\bm \epsilon} {\bf E}).{\bf E} 
        = -\frac{1}{2}\epsilon_0 {\bm \epsilon} \nabla U. \nabla U,
\end{equation}
where \textit{U} is the electric potential and ${\bf E} = −\nabla{U}$.

Using the commonly assumed one-constant approximation for the elastic constants of the NLC \cite{mottram2014introduction}, the elastic  energy density and thermotropic energy density may be written as: 
\begin{equation}
F_{el}=\frac{L}{2} |\nabla {\bf Q}|^2\quad\quad \mbox{and}\quad\quad F_{th} = A {\rm tr}(\mathbf{Q}^2) + \frac{2B}{3}{\rm tr}(\mathbf{Q}^3) + \frac{C}{2}{\rm tr}\left(\mathbf{Q}^4\right), 
\end{equation}
where $L$ is an average elastic constant related to the splay bend and twist constants $K_{11}$, $K_{22}$ and $K_{33}$, ${\rm tr}(.)$ signifies the trace of a tensor,  $A = a(T − T^\ast) = a\Delta T$ is a temperature dependent coefficient, with $a > 0$ and $T^\ast$ the temperature at which the isotropic state ${\bf Q}=0$ becomes unstable, and $B$ and $C$ are assumed to be temperature independent. In the simulations below, these coefficients were set to $A = −6.5 \times 10^5 Nm^{-2}$, $B = −16 \times 10^5 Nm^{−2}$ and $C = 39 \times 10^5 Nm^{−2}$ \cite{mottram2014introduction}.

The total free energy density is therefore written as:
\begin{equation}
\label{freeenergy}
{\cal F} = \int_V{\frac{L}{2} |\nabla{\bf Q}|^2 + A {\rm tr}(\mathbf{Q}^2) + \frac{2B}{3}{\rm tr}(\mathbf{Q}^3) + \frac{C}{2}{\rm tr}\left(\mathbf{Q}\right)^4 − \frac{1}{2}\epsilon_0 \epsilon \nabla U. \nabla U} \ dV. 
\end{equation}

The equilibrium director configuration can be obtained by minimizing the total free energy in eq.~\ref{freeenergy} using the calculus of variations, which leads to the partial differential equations for the minimizing ${\bf Q}$-tensor components, 
\begin{align}
\label{ELeq}
    \left( \,\frac{\partial F_b}{\partial Q_{i,j}} \right )_{,j} \, &= \frac{\partial{F_b}}{\partial Q_i},
    \end{align}
where the $_{,j}$ subscript denotes partial differentiation with respect to the ${\rm j}^{\rm th}$ component of the position vector, which in our present 2-dimensional setting is ${\bf x}=(x,z)$, and the summation convention has been used.
Together with the equation for the components of the ${\bf Q}$-tensor, Gauss's law for the electric potential must be simultaneously solved, 
\begin{align}
\label{GLaw}
    0=\nabla .{\bf D}=\nabla .\left(\epsilon_0 ({\bm \epsilon} {\bf E})\right)=-\epsilon_0\nabla .\left( ({\bm \epsilon} \nabla U)\right)=-\epsilon_0\left( { \epsilon}_{ij} U_{,j}\right)_{,i}.
    \end{align}
    
In our previous work, a 2x2 Jones matrix method was used to calculate the optical properties for a specific direction profile ${\bf n}({\bf x})$ \cite{ghannam2021reconfigurable}, where the optical properties of the cell is  calculated using a multi-layer method, dividing the LC cell into thin layers, assuming the director to be constant in each layer, and thus determining the transmittance through each layer. This method is suitable for thin cells since the reflections between the layers are neglected. However, in the current tunable antenna design, a relatively thick LC cell ($80\,\mu$m) is used, and the Berreman 4x4 matrix method, a more accurate optical model, is introduced here to calculate the transmittance. This optical calculation includes consideration of multiple reflections of the electric and magnetic waves between layers. The dielectric permittivity is then a function related to the axis $z$ perpendicular to the layers and the Berreman matrix, which encodes the optical electric and magnetic fields within each layer, can be written as:
\begin{equation}
\setlength{\arraycolsep}{10pt}
{\cal D}=\left[                 
  \begin{array}{cccc}   
    -\eta\dfrac{\epsilon_{31}}{\epsilon_{33}} & 1-\dfrac{\eta^{2}}{\epsilon_{33}} & -\eta\dfrac{\epsilon_{32}}{\epsilon_{33}} & 0\\[5mm]  
    \epsilon_{11}-\dfrac{\epsilon_{13}\epsilon_{31}}{\epsilon_{33}} & -\eta\dfrac{\epsilon_{13}}{\epsilon_{33}} & \epsilon_{12}-\dfrac{\epsilon_{13}\epsilon_{32}}{\epsilon_{33}} & 0\\[5mm]  
    0 & 0 & 0 & 1\\[2mm]  
    \epsilon_{21}-\dfrac{\epsilon_{23}\epsilon_{31}}{\epsilon_{33}} & -\eta\dfrac{\epsilon_{23}}{\epsilon_{33}} & \epsilon_{22}-\dfrac{\epsilon_{23}\epsilon_{32}}{\epsilon_{33}}-\eta^{2} & 0\\ 
  \end{array}
\right],     
\end{equation}
where $\eta$ is the component of the incident light in the $x$-direction.

Due to the symmetry of the permittivity tensor, ${\cal D}_{13}={\cal D}_{24}$, ${\cal D}_{11}={\cal D}_{22}$ and ${\cal D}_{41}$=${\cal D}_{23}$, which allows the Maxwell equations to be simplified and the output light  vector to be determined \cite{yang2014fundamentals}. This modelling method was used to calculate the transmittance of the LC cell.\\
\vspace{5mm}


\subsection{Design}
To realize the controllable antenna, we proposed a multi-layer substrate with a sandwich structure consisting: dielectric layers, grid electrodes, LC material and alignment layers. The dielectric layer, typically polyimide (PI), protects the user's eyes from the electric field. A transparent electrode layer consisting of indium tin oxide (ITO) was used, as well as a planar aligment layer, which can be achieved by rubbing the PI layer in one direction, creating microgrooves along which the LC aligns \cite{toney1995near}. Moreover, the LC cell thickness can be fixed through the use of  SiO$_{2}$ micro-spheres spacer within the LC layer. 

In Table \ref{tab:LCmaterials}, we list four commonly available NLC materials and mixtures. When choosing the NLC material for this application low LC driving voltages are needed to ensure device safety and portability, and therefore LCs with high $\Delta \epsilon$ and low $K_{11}$ are more suitable for low-power applications. Furthermore, relatively low splay, twist and bend elastic constants ($K_{11}$, $K_{22}$ and $K_{33}$) are required to reduce the device response time. Based on these considerations, RDP-84909 (DIC cooperation) was chosen as the most suitable LC material. 

\begin{table}[htb]
\centering
 \caption{Material parameters for the LCs considered during this study \cite{coles1978laser, miroshnychenko2005evolution, cummins1975dielectric, blinov1996electrooptic, wang2004correlations,podoliak2012magneto,al2017temperature}. The parameters for RDP-84909 are provided by DIC corporation, Japan.}
 \begin{threeparttable}
  \begin{tabular}[htbp]{p{5cm}|p{2.5cm}|p{2.5cm}|p{2.5cm}|p{2.5cm} }
    \hline
    Parameter & 5CB & BLO48 & RDP-84909 & E7\\
    \hline
    Clearing temperature, $T^{NI}(^{\circ}{\rm C})$ &  35       &  100    &    91.7     &    59.85  \\
    $\epsilon_\perp$                            &  6.9      &  5.2    &    8        &    5.17   \\
    $\Delta \epsilon$                           &  11       &  16.9   &    39.1     &    14.37  \\
    $n_o\ (\lambda = 532\,{\rm nm}) $                   &  1.532    &  1.573  &    1.48     &    1.53   \\
    $ \Delta n\ (\lambda = 532\,{\rm nm})$              &  0.174    &  0.226  &    0.1464   &    0.22   \\
    $ K_{11}\ ({\rm pN})$                              &  6.4      &  15.5   &    4.6      &    11.2   \\
    $ K_{22}\ ({\rm pN})$                              &  3        &  12     &    1.2      &    6.8    \\
    $ K_{33}\ ({\rm pN})$                              &  10       &  28     &    13.8     &    17.8   \\
    \hline
  \end{tabular}
  \end{threeparttable}
  \label{tab:LCmaterials}
\end{table}

A spiral antenna was chosen since it is a commonly used antenna geometry in wearable applications, due to its simplicity, universality and low-cost \cite{baghernia2020broadband,egorov2000non}. Our antenna was designed for 5G applications, where we chose n78 from frequency range 1 (FR1) of the 5G mobile network, i.e., $3.3$–$3.8\,$GHz \cite{access2011user}. The electromagnetic waves in this range are defined as centimeter waves. Due to their small wavelength, centimeter waves could be received in narrow beams, which is suitable for data linking, point to point communication and radar applications.

As is shown in Fig.~\ref{Schematic_Diagram}(b), a single-armed Archimedean spiral coil was chosen for its typical wide bandwidth and circular polarization, and impedance matching was used due to the geometry limitation. A matching parallel capacitor was used to adjust the antenna to the desired resonant frequency with maximized power delivery, using the equation \cite{pozar2011microwave}:
$$f=\frac{1}{2\pi\sqrt{L_{s}C_{p}}},$$
where $L_{s}$ is the equivalent inductance of the spiral antenna and $C_{p}$ represents the total capacitance of the device. Finite element analysis was carried out with High Frequency Structure Simulator (HFSS) software, where the Archimedean spiral geometry is defined as: $r=r_{\theta}+\alpha\phi$, where $r_{\theta}=r_{0}$ and $r_{\theta}=r_{1}$ are the inner and outer radius of the starting circle respectively, $\alpha$ is the growth rate of spiral arm,  $\phi$ denotes the angle or the arc value, and the  spiral lines were defined as $x_{t}=r_{\theta}e^{\alpha t}\cos{t}$, and $y_{t}=r_{\theta}e^{\alpha t}\sin{t}$. The specific parameters of antenna were defined as follows: $\alpha = 17.5\,$mm, $\phi = 6\pi$, $r_{0} = 4.5\,$mm, $r_{1} = 4.7\,$mm, $C = 0.8\,$pF. Copper elements define the inductance of the spiral coil and the gaps between generate the equivalent capacitance of the device.

\begin{figure}[H]
  \centering
  \includegraphics[width=18cm]{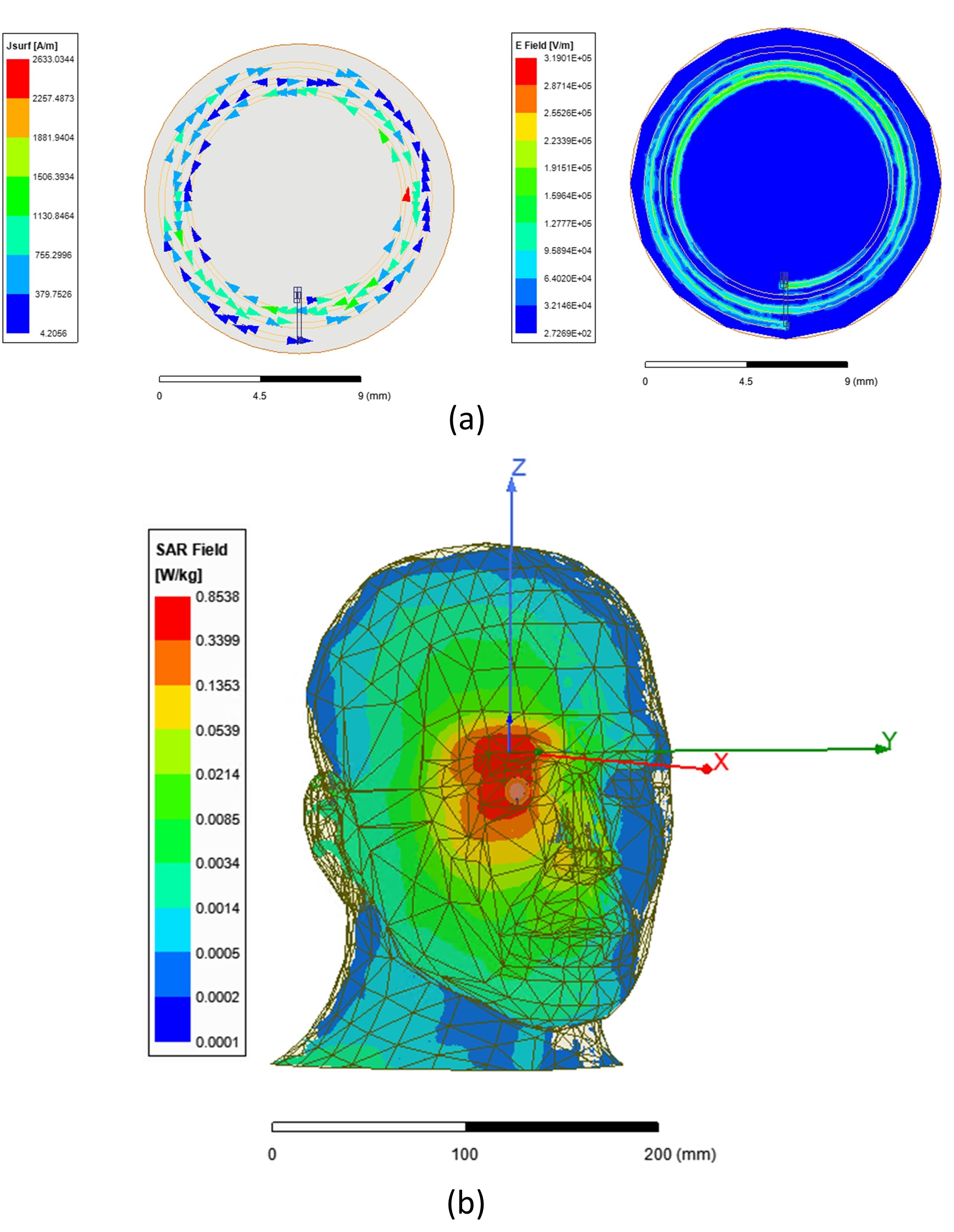}
  \caption{(a) The surface current density on the antenna and the electric field intensity on the top surface when the LC director is planar aligned ($V=0$). (b) Specific Absorption Rate (SAR) simulation, demonstrating that the antenna is safe for near-eye operation.}
  \label{surfcurrent}
\end{figure}

The surface current distribution presented in Fig.~\ref{surfcurrent}(a) indicates the current flows on the antenna surface, which is restricted by the spiral geometry. Additionally, the induced electric field indicates a higher concentrated electric field intensity on the intermediate spirals. The Specific Absorption Rate (SAR) simulated was constrained to $0.8538\,$W/kg as is shown in Fig.~\ref{surfcurrent}(b), which is within the safety regulation standard ($2\,$W/kg). \cite{lin2006new}.

\begin{figure}[H]
  \centering
  \includegraphics[width=18cm]{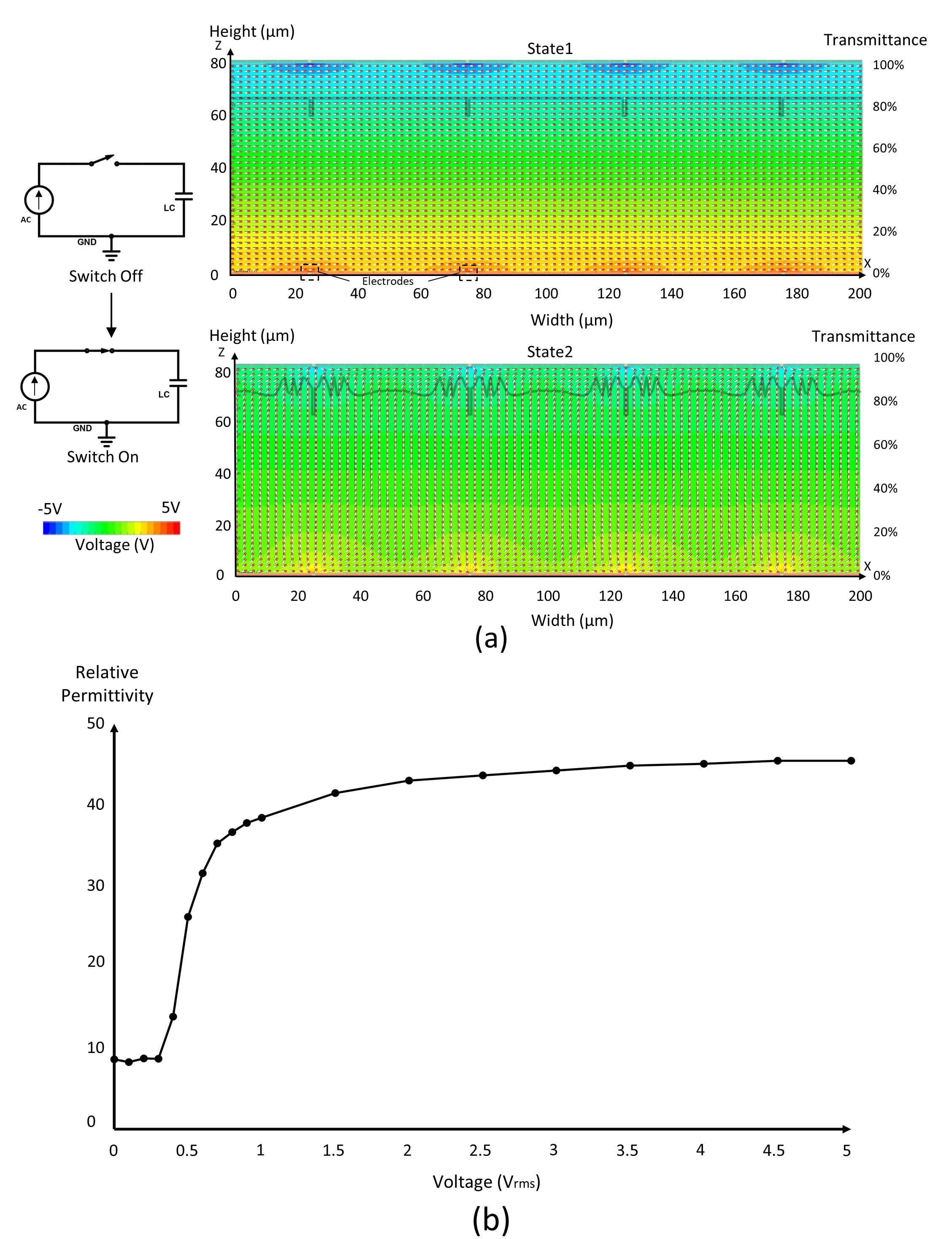}
  \caption{\textcolor{red}{(a) The electrode structure and ${\bf Q}$-tensor simulation results of the liquid crystal layer. The LC substrate can be switched from state 1 to state 2. (b) Plot of the Freedericksz curve for RDP-84909.}}
  \label{LC}
\end{figure}

Grid electrodes were chosen on both the top and bottom electrode layers rather than plate electrodes since the latter would reflect electromagnetic wave and prevent it from reaching the reconfigurable LC layer. In Fig.~\ref{LC}(a), a top-bottom grid electrode array with $1\,\mu$m width and $49\, \mu$m gap (2\% of grid density) were simulated using the ${\bf Q}$-tensor theory described above. The grid electrodes have similar performance (in terms of electric field intensity and direction) to plate electrodes and, consequently, the LC molecules exhibit similar behavior with each type of electrode. The transmittance of the cell was then calculated using the 4x4 Berreman matrix method described above, and is also shown in Fig.~\ref{LC}(a). The transmittance through out the cell is between 91\% to 74\%. The lowest transmittance occurs where the electrodes are located and therefore the LC cell with grid electrodes would have higher average transmittance than plane electrodes.

The threshold voltage required to tilt the LC directors from planar towards the vertical  can be determined using \cite{yang2014fundamentals}$$V_{c}=\pi\sqrt{\frac{K_{11}}{\epsilon _{0}\Delta \epsilon}},$$ which shows that the cell thickness does not affect the threshold voltage. Although a thick cell will result in a relatively long response time \cite{kim2021design}, thicker LC cells will have a more significant frequency bandwidth variation from  switching the director configuration of the LC substrate. In this case, thicker substrates lead to wider antenna tuning frequencies. Therefore, to ensure both the high tunability and low response time, the thicknesses of the LC and PI layers were set to $80\,\mu$m and $50\,\mu $m, respectively.

\section{Results and Discussions}
\begin{figure}[H]
  \centering
  \includegraphics[width=15cm]{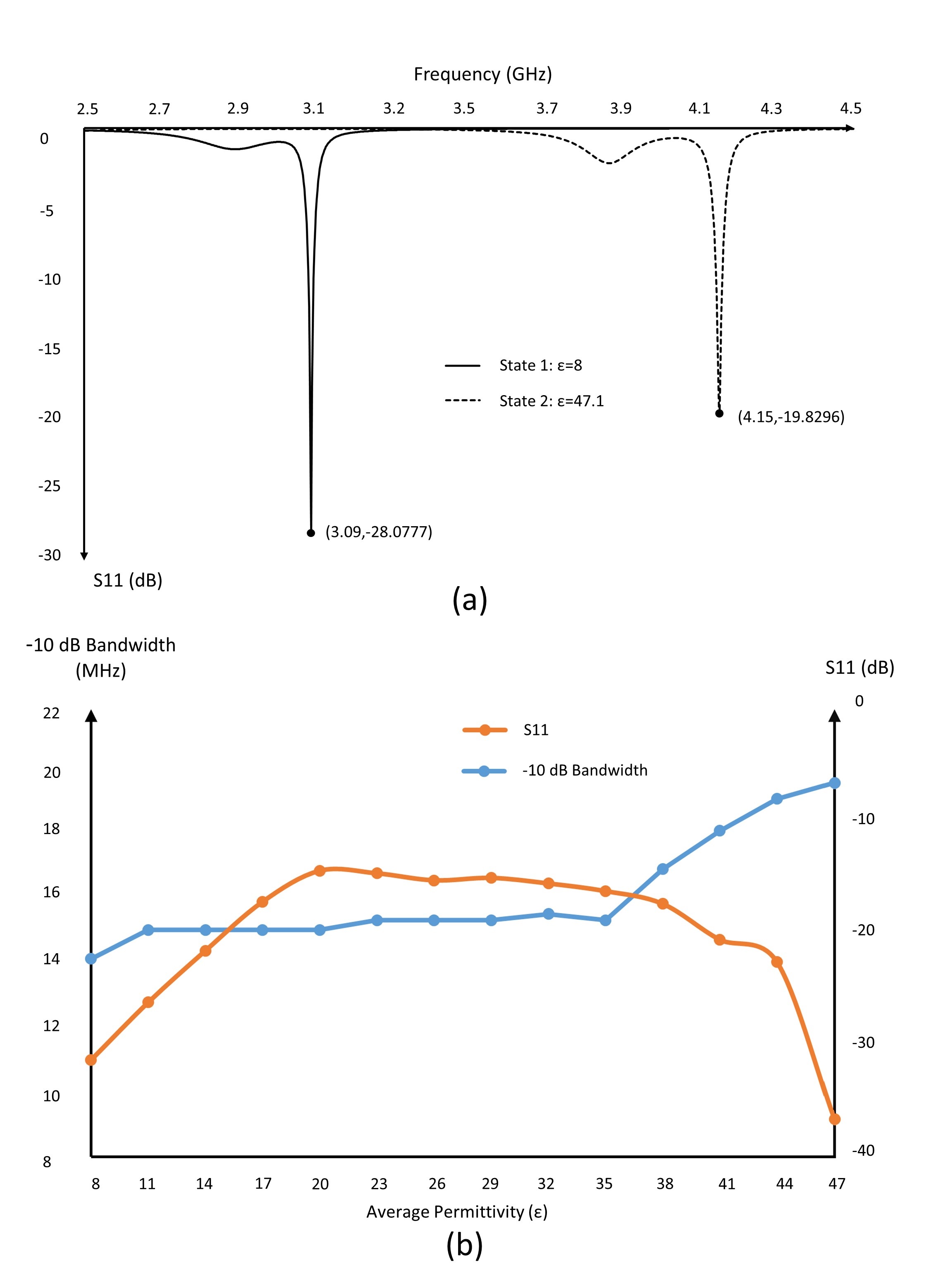}
  \caption{\textcolor{red}{(a) Plots of the S11 parameter of the antenna when the substrate is in two different states. This figure shows the theoretical maximum tunability. (b) Plots of S11 and -10 dB bandwidth when the permittivity of the LC layer varies from $8$ to $47$. Therefore, the antenna could operate with the proposed variation of substrate permittivity.}}
  \label{Director_profile}
\end{figure}

As mentioned previously, to reduce the response time, the antenna was mounted on a 180 $\mu$m reconfigurable substrate. Assuming a relatively weak  anchoring effect, we can assume that all LC directors are planar and homeotropically aligned in state 1 (no applied voltage) and state 2 (with applied voltage), respectively. In this case, the theoretical maximum tunable frequency range can be calculated. For the liquid crystal RDP-84909, the permittivities of the two states are 47.1 and 8, respectively, and therefore the maximum frequency shift is from $3.09$ to $4.15\,$GHz (Fig.~\ref{Director_profile}(a)), which is in the centimeter wave range.

In practice, due to the anchoring effect of the PI layer on the LC, the directors in contact the edge of the alignment layers tend to maintain their original orientation. Thus, when a voltage larger than the threshold voltage is applied to the cell, the average permittivity of the cell will be smaller than $\epsilon_\parallel$. In future works, the average capacitance between electrodes could be calculated using the simulation software TechWiz. Hence, the relative permittivity could be calculated with various input voltages. However, using the model in this work, in Fig.~\ref{LC}(b) the average permittivity is plotted against voltage, known as the Freedericksz curve, and shows a sharp rise of relative permittivity around $0.35\,$V, which is in agreement with the theoretical threshold voltage of RDP-84909 calculated as $0.362\,$V. Moreover, the permittivity increases significantly between $0.35$ and $1\,$V, where the permittivity varies from $8$ to $38.4$. The relative increment of permittivity decreases after $2\,$V. According to the Freedericksz curve, we see that, for significant changes to the permittivity it is possible to operate the reconfigurable substrate within a relatively small voltage range, between $0.35$ to $1\,$V is preferred, an ideal range for wearable applications.

 Simulations were then performed to obtain the relationship between LC substrate permittivity found through the Freedericksz curve, and the resonant frequency. In the antenna model, the substrate contains a dielectric layer and an LC layer, where we here  set the permittivity of the LC layer as variable, and thus obtain the frequency shift in response to this variation of permittivity. As is shown in Fig.~\ref{Director_profile}(a), when the average permittivity of the substrate is increased from $8$ to $47.1$ (we neglect the anchoring effect to achieve maximum permittivity), the resonant frequency is shifted from $3.09$ to $4.15\,$GHz, successfully covering the target frequency range n78 ($3.3$ to $3.8\,$GHz). The corresponding average permittivity varies from $14$ to $32$. Moreover, we also obtained -10dB bandwidth versus different permittivity, as is shown in Fig.~\ref{Director_profile}(b). This curve illustrate that the -10dB bandwidths are within the range of $14.2$ to $19.7\,$MHz, which can ensure the data transmission is taking place efficiency and at an acceptable speed.

\begin{figure}[H]
  \centering
  \includegraphics[width=15cm]{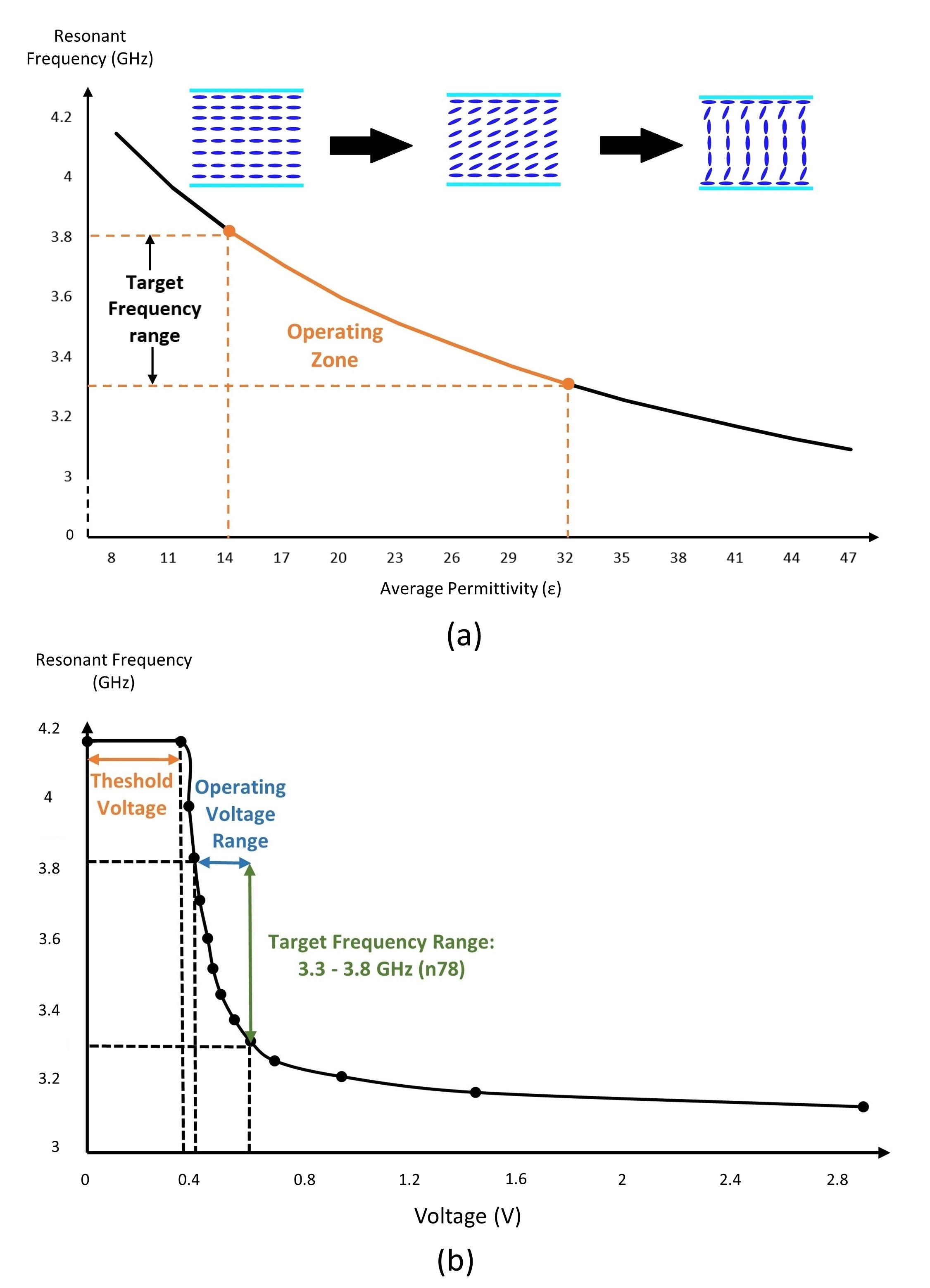}
  \caption{\textcolor{red}{(a) Average permittivity versus resonant frequency. This plot demonstrates that the antenna could cover the target frequency range (n78: $3.3$-$3.8\,$GHz). (b) Plot of   resonant frequency versus  applied voltage. The driving voltage required to cover the target frequency  ranges from $0.4$ to $0.6\,$V.}}
  \label{Tunability}
\end{figure}

Combining the Freederickz curve and the relationship between applied voltage and resonant frequency, shown in Fig.~\ref{Tunability}(a) we obtain the relationship between the applied voltage and the resonant frequency, as shown in Fig.~\ref{Tunability}(b). This figure plots the applied voltage from $0$ to $3\,$V, since the antenna is designed for wearable application, and shows that when the voltage is varied the resonant frequency varies from $3.09$ to $4.15\,$GHz. As mentioned previously, the target frequency range of this antenna design is from $3.3$ to $3.8\,$GHz which is covered when the voltage is between $0.4$ and  $0.6\,$V, which is suitable for low power consumption wearable applications.

In Table 2, the specifications of our antenna are compared with previous tunable antenna designs. The table lists important parameters for assessing the performance of the tunable antenna including antenna type, area, frequency range, driving voltage and substrate thickness. It is clear that our tunable antenna design has a smaller footprint, thinner substrate material and lower driving voltage compared to other designs, and could potentially be used for wearable or implantable applications.

\begin{table}[ht]
\centering
 \caption{Comparisons with previous work}
 \begin{threeparttable}
  \begin{tabular}[htbp]{p{1.7cm}|p{2.5cm}|p{2cm}|p{3cm}|p{2cm}|p{2.5cm} }
    \hline
    Reference & Antenna Type & Antenna Area (mm$^{2}$) & Frequency Range (GHz)  & Driving Voltage (V) & Thickness of Substrate ($\mu$m)\\
    \hline

    \cite{xu2017differential} &  Patch Antenna      &  1026.4    &    2.532-2.427    &    25  & 787\\
    
    \cite{sun2020electronically} &  Patch Antenna     &   0.225   &    85-115         &   20   & 1020.2\\
    
    \cite{kim2021design} &   Patch Antenna      &   400   &    4.84-5.12    &   -   & 300 \\
    \cite{missaoui2019electrically} &   Patch Antenna      &   612   &    2.02-2.4   &   -   & 500\\
    \cite{martin2003patch} &   Patch Antenna      &   434.7   &    2.7-2.9   &   -   & 500\\
    
    This Work        &  Spiral Antenna      &  6.8   &    3.09-4.15        &    0.6  &  180 \\
    
          &        &     &    Target: 3.3 to 3.8      &      &  \\
    \hline
  \end{tabular}
  \end{threeparttable}
  \label{materials}
\end{table}

\section{Conclusion}
In this paper, we have proposed a tunable centimeter wave antenna design for smart glasses applications. The antenna could be tuned from $3.3$ to $3.8\,$GHz (FR1 n78 for 5G communication) using $0.4$ to $0.6\,$V. The theoretical maximum tunability of this design is $1.06\,$GHz ($3.09$-$4.15\,$GHz), which could be achieved only when the anchoring effect is neglected. Compared with previous designs, this antenna has lower driving voltage, smaller area and high tunability. This prospective design is suitable for wearable and implantable devices. The low driving voltage could allow this tunable antenna to be used in low-consumption and autonomous devices. The small area of the antenna could also enable its integration within a miniaturised design and its high tunability means that it can cover more frequencies and communicate with a wider range of devices compared to currently available designs.

Further work will focus on fabricating the antenna and developing the driving electronics for the reconfigurable surface. To fabricate the antenna pattern, laser printing or lithography techniques can be used, and glass spherical spacers would be used to control the thickness of the liquid crystal cell. The structure of the antenna and LC material will also be optimized to cover a wider bandwidth. Following this, the intention is to embed this antenna within a smart glasses system to expand the bandwidth allowing communication with multiple external devices, which will be of considerable benefit for IoT applications.  

\bibliographystyle{MSP}
\bibliography{Reference}

\end{document}